\documentclass{elsart}
\usepackage{graphicx,natbib,amssymb} 
\journal{Physica A}
\begin{document}
\begin{frontmatter}

\title{
LOCALLY PREFERRED STRUCTURE AND FRUSTRATION IN GLASSFORMING LIQUIDS: A
CLUE TO POLYAMORPHISM?}
\author[1]{G. Tarjus},
\author[2]{ C. Alba-Simionesco},\author[1,3]{ M. Grousson},
\author[1]{ P. Viot} and \author[3]{ D. Kivelson}
\address[1]{ Laboratoire de Physique Theorique des Liquides,
Universit{\'e}  Pierre  et  Marie  Curie,     4   Place Jussieu,     Paris
75005, France}
\address[2]{ Laboratoire de Chimie Physique, b{\^a}t 349, Universit{\'e}
 Paris Sud, Orsay 91405, France}
\address[3]{ Department of Chemistry
 and  Biochemistry, University of California,  Los  Angeles, CA 90095,
 USA}

\begin{abstract}
We propose    that the concept   of liquids  characterized by  a given
locally preferred   structure (LPS) could   help  in understanding the
observed  phenomenon  of  polyamorphism.  ``True polyamorphism'' would
involve  the   competition between  two  (or  more) distinct  LPS, one
favored at low pressure  because of its low energy  and one favored at
high pressure   because   of   its  small  specific  volume,    as  in
tetrahedrally coordinated systems. ``Apparent polyamorphism'' could be
associated with the existence  of a poorly crystallized defect-ordered
phase  with a  large unit cell  and small  crystallites, which may  be
illustrated by the metastable glacial phase of the fragile glassformer
triphenylphosphite;  the apparent   polyamorphism  might  result  from
structural frustration, i. e., a  competition between the tendency  to
extend the  LPS  and a global  constraint that  prevents tiling of the
whole space by the LPS.
\end{abstract}
\end{frontmatter}

\maketitle

\section{Locally preferred structures in liquids}
Polyamorphism  is the coexistence  of  condensed  phases of  identical
chemical composition  that appear amorphous,  i.  e.,  with no obvious
long-range order.  This is to be  distinguished from  those situations
such as concentration-driven transitions in multi-component liquids or
from  gas-liquid coexistence.   This puzzling   phenomenon, not to  be
confused  with the long  recognized   polymorphism between phases   of
different  symmetries   (be they  crystals,  liquid  crystals, plastic
crystals,     etc.),        has        recently     attracted     much
attention\cite{mish1,ang1,mcmil,kiv,kata,yarg,lacks,muk}.  It  has
been observed in  liquids at low temperature,  usually in the vicinity
of the glass transition.  In this article, we suggest that the concept
of   {\it locally  preferred structure}   in   liquids is central   to
understanding polyamorphism.   A locally  preferred  structure can  be
loosely  defined as an arrangement of  molecules which, in a given
region of the pressure-temperature phase diagram, minimizes some local
free energy.

Most recorded  examples of polyamorphism are tetrahedrally coordinated
systems,    such     as       H$_2$O,   Si0$_2$,        or     Ge0$_2$
\cite{mish1,ang1,yarg,lacks,muk,mish2},    in  which   low-temperature
coexistence of  amorphous phases is  observed  under sufficiently high
pressure.  This phenomenon  can be rationalized   in terms  of  {\it a
competition between different locally  preferred structures}.  This is
best illustrated  by   the  case of  water,  a  system  that has  been
thoroughly   studied   by     Gene    Stanley   and    his   coworkers
\cite{mish2,stan}.

Following the picture  put  forward by Stanley and   coworkers, liquid
water  has been characterized   schematically by two kinds of  locally
preferred  structures (LPS) that  are favored  in different regions of
the  pressure-temperature phase    diagram.   This  can be   seen   by
considering the  arrangements of $5$-molecule  clusters, also known as
Walrafen pentamers \cite{mish2,stan}.   In  one  LPS,  two neighboring
pentamers are  oriented  relative to   each other so   as  to form two
linking hydrogen bonds; this is a low-energy, but open (large specific
volume) structure, and due to the directional nature of the H-bonding,
it has a low entropy. In the other arrangement, the two pentamers come
closer to each other, but are no longer  bonded: the structure is then
better packed (small specific volume) and has a higher entropy, but at
the expense  of a higher   energy.   The former  structure is  locally
favored  at low  pressure whereas the   latter is locally preferred at
high  pressure.   As it   has been shown  via   a description   of the
intermolecular interactions  in  terms of  an  effective,  spherically
symmetric pair potential \cite{sadr}, the competition between two such
LPS's may  lead  to  a {\it  bona   fide} phase transition   between a
low-density    liquid  and     a    high-density   liquid    in    the
pressure-temperature diagram.   Similar   reasoning   could apply   to
glassforming  liquids  such  as SiO$_2$   and GeO$_2$ whose  LPS  is a
$4$-coordinated cation at low  pressure  and a $6$-coordinated one  at
high pressure \cite{yarg}.  Another  extreme example is also  provided
by  the polyamorphism  of liquid  phosphorus,  one  liquid phase being
characterized  by a local  organization formed  by $P_4$ molecules and
the other   one being  a  polymeric-like  phase  of phosphorus atoms
\cite{kata}.

We suggest  here that in some systems,  in addition to the alternative
LPS's,  there can also be a  competition associated with the inability
of   a  given  LPS   to  tile   space,  i.e.,  with    {\it structural
frustration}. In this  case the polyamorphism  may incorporate  a very
poorly developed  mesoscopic  order   and  so might  be   described as
"apparent polyamorphism". Such  "apparent polyamorphism" appears to be
illustrated by triphenylphosphite (TPP).

\begin{figure}
\begin{center}
\resizebox{15cm}{!}{\includegraphics{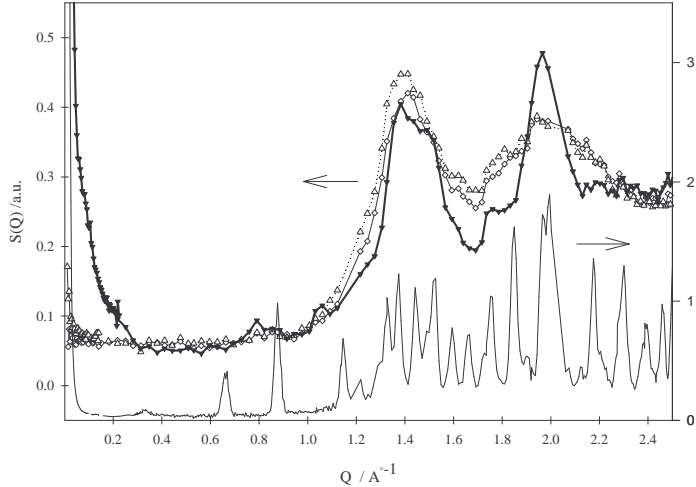}}
\caption{Combined low- and wide-angle neutron scattering data
 (in arbitrary units) for the static structure factor $S(Q)$ of TPP in
 its different phases: supercooled liquid (open triangles, $T=218 K$),
 glass (open diamonds,  $T=183-184 K$), crystal (thin continuous line,
 $T=270  K$  and $183  K$  at  low  $Q$), and   glacial  phase (filled
 triangles, $T=225 K$) phases.  For  clarity the $S(Q)$ of the crystal
 is shifted downwards. The  melting temperature is  $295 K$, the glass
 transition temperature  of the supercooled  liquid is around $195 K$,
 and the liquid-glacial transition temperature is around $240 K.$}

\end{center}
\end{figure}
\section{Apparent polyamorphism in TPP}
Triphenylphosphite (TPP) is one of    the most fragile   glassformers,
i. e., one  for which  the increase  of viscosity and  relaxation time
with  decreasing temperature is   most  dramatic.  A new metastable
phase,  denoted ``glacial  phase'',    has  recently been  observed   at
atmospheric pressure \cite{kiv}. This phase transforms to and from the
supercooled liquid and is    metastable  with respect to the    normal
crystal \cite{kiv,demirj,van,mizu}.

\begin{figure}
\begin{center}
\resizebox{15cm}{!}{\includegraphics{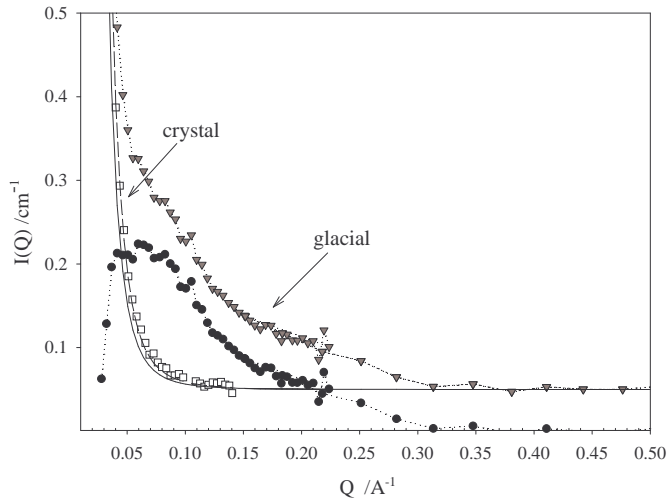}}
\caption{
 Low-$Q$ region for the crystal (open squares) and the glacial
 phase (filled triangles): it shows the $Q^{-4}$
 Porod's regime for the crystal (full line) and the glacial phase
 (dashed line), as well as the distinct shoulder in the glacial-phase
 data; this latter can be interpreted as the superposition of
 a Porod's contribution and a broad peak (filled circles) that
 is indicative of structural organization on a mesoscopic scale
 (see also \cite{alba}).}
\end{center}
\end{figure}
Since its discovery, the glacial phase of TPP has been studied by many
groups and by means of a variety of experimental techniques
\cite{kiv,demirj,mizu,joh,ross,desc,alba,senk,schwi}.  A number of
conjectures have been proposed concerning the structure of the glacial
phase,  and most of them describe  the phase as amorphous.  The reason
for this is that in normal  X-ray or neutron scattering, the structure
of the glacial phase does not show well defined Bragg peaks as 
observed in crystalline materials. This is illustrated in Fig.~1, where
we display the result of  a series  of neutron scattering  experiments
carried out  both on the $D7$ spectrometer  of the ILL in Grenoble (in
the range  of wave-vector $Q$ from $0.1$  to  $2.5$ {\AA}$^{-1}$) and on the
small-angle spectrometer PAXE  at the LLB in  Saclay (in the $Q$-range
between   $0.01$ and  $0.12$ {\AA}$^{-1}$)\cite{alba}.  The static structure
factor $S(Q)$ of the   glacial phase  is   distinct from that of   the
liquid,  the glass,  and  the  crystal,  and, although  the  peaks are
somewhat sharper than those in the liquid and the glass, they are much
broader  than     those   of the  crystal.      However, what  clearly
distinguishes the structure factor of the  glacial phase is an unusual
feature at small $Q$'s, a feature that is visible in the experiment on
D7 but shows  up more clearly  in the small-angle  scattering data: in
sharp  contrast with the  $S(Q)$'s of the  other phases, the $S(Q)$ of
the glacial phase has  a pronounced shoulder for $Q<0.2$ {\AA}$^{-1}$,  and,
in addition, the scattered intensity  keeps rising very steeply at the
lowest $Q$'s in a manner that is  compatible with the $Q^{-4}$ Porod's
law    observed  in  powders      of   crystalline   materials:    see
Fig. 2

By analyzing  the low-Q scattering data  described above (and shown in
Fig.   2) as the superposition  of a Porod's tail  and of a broad peak
centered at $Q_P\simeq 0.07$ {\AA}$^{-1}$, standard  arguments used in studying
polycrystals indicate that the  ``apparently amorphous'' glacial phase
could be a powder of an  unusual crystalline material characterized by
a  large unit cell of typical  size ($2\pi /Q_P=   80$ {\AA}) and with small
polydisperse crystallites of about  $100$ to $250$ {\AA}\cite{alba}:  this
is   sketched in  Fig.~3.   The   premelting  phenomenon reported   in
Ref.\cite{demirj} is   also   consistent   with this   picture  of   a
crystalline structure with small  crystallites.   In such a   ``poorly
crystallized''  material, the small number of  unit cells contained in
the crystallites,  the   polydispersity   of the  crystallites,    the
rotational disorder,  and strain effects could  all combine to explain
the absence of well defined Bragg peaks, thus resulting in a structure
that at first glance looks amorphous.

\begin{figure}
\begin{center}
\resizebox{12cm}{!}{\includegraphics[angle=270]{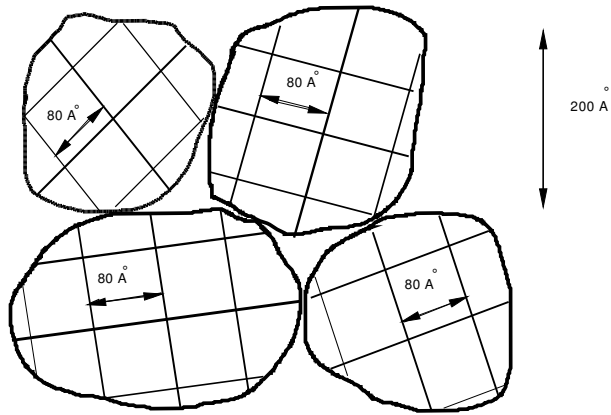}}
\caption{ Speculative picture of the structure of the glacial phase
 of TPP:  a   poorly crystallized material   with  a large  unit  cell
 ($80${\AA}),  small polydisperse  crystallites   (typically, $200${\AA}), and
 interstitial liquid.}
\end{center}
\end{figure}

What then is  the physical origin of  this "apparent" polyamorphism at
atmospheric pressure?

\section{Structural frustration}

If one accepts  the premise that there exists  a LPS in a liquid (say,
at  atmospheric pressure so  that   competition between two  different
LPS's  is unlikely),  one must  worry   about a  competition between a
tendency  to extend  the  LPS and a  global  constraint.  It has  been
suggested that {\it  structural frustration}, i. e., the impossibility
to tile the whole space by periodically  replicating the LPS, might be
ubiquitous in  glassforming liquids \cite{sadoc,nelson,kivelson}.  One
manifestation of frustration can be perceived in the fact that liquids
restructure and undergo  a strong first-order  transition to a crystal
whose local  structure is  different   from the  LPS; this  transition
occurs in order  to avert the increasing  strain that develops as  the
temperature decreases and the LPS grows. The frustration may also play
a dominant role in glass formation.
	
The canonical example  of  structural   frustration is  provided    by
single-component systems of spherical particles interacting via simple
pair potentials,  and the phenomenon is  best illustrated by comparing
the situations  encountered in $2$     and $3$  dimensions.  In    $2$
dimensions, the arrangement  of disks that  is locally preferred is  a
hexagon  of six  disks  around  a central    one, and  this  hexagonal
structure  can be  extended to the  whole space  to form a  triangular
lattice.  In  $3$ dimensions,  as    was shown   long  ago by    Frank
\cite{frank},   the locally preferred     structure of spheres  is  an
icosahedron, but the $5$-fold rotational  symmetry of the icosahedron is
incompatible with translational   periodicity,  and  formation  of  an
icosahedral   crystal is precluded. Frustration  is   thus absent in $2$
dimensions, but present in $3$ dimensions.  As a result, crystallization
is essentially continuous in the former case, and neither supercooling
of the  liquid nor glass formation   are possible. On  the other hand,
crystallization  of spheres in $3$  dimensions is a strongly first-order
transition that  involves a restructuring of  the  local order to form
the face-centered-cubic (or the hexagonal-close-packed) order that can
tile space periodically. Studies of structural frustration for spheres
in  $3$  dimensions have  been  further developed   to describe metallic
glasses \cite{sadoc,nelson,sachdev}.
	
Two points are worth stressing: first,  frustration can be relieved by
"curving" the regular 3-dimensional  Euclidean space, so that an ideal
world without  frustration is generated  where periodic  tiling by the
LPS  (e.g., icosahedral  order) is  possible. Forcing  the ideal order
into the real world  leads to the  formation of  defects (disclination
lines in the example  of spherical particles  in 3 dimensions)  and to
the growth of  a strain free-energy  that opposes the extension of the
LPS. Secondly, back in the Euclidean space ordered phases can still be
formed which are  different from the usual and  more stable crystal in
that their structure is partly based on the LPS. The system can indeed
get around frustration and form ``defect ordered phases'' in which the
defects themselves form periodic  structures with long-range order, as
in  the Frank-Kasper  phases   \cite{sadoc,nelson}.  In real  metallic
systems such phases are  only observed in alloys  made of two or  more
components, but  a  recent   simulation   study   has shown  that    a
one-component  atomic liquid     whose  particles  interact    with  a
spherically symmetric  potential that  favors local  icosahedral order
can form a  metastable 'defect-ordered  phase'\cite{dzug}; this latter
is  a dodecagonal  quasi-crystal that is  essentially  a layered phase
with  translational periodicity  in one  direction but quasi-periodic,
icosahedral-like order in the transverse directions.

The example  of spheres and their  local icosahedral  packing symmetry
has been introduced only   for illustration.  More generally, one  can
envisage competition  between the tendency to  extend the  LPS and the
global constraint embodied    in the  structural   frustration  as  an
intrinsic feature of all liquids.  A coarse-grained description should
thus be sufficient, and it has been argued  that a minimal model could
be built,   based on competing  effective  interactions acting on very
different  length scales   \cite{kivelson}:   a  short-ranged ordering
interaction, that  describes the tendency  to extend the LPS and leads
to a continuous   transition to an ideal  crystal  in the  absence  of
frustration, and a weak but  long-ranged (1/r) frustrating interaction
that generates  a super-extensive   strain  free-energy  opposing  the
growth of the ideal structure. It has been recently  shown by means of
model-calculations that such  ingredients do indeed  lead to a  strong
slowing   down of the relaxations  as  the temperature is lowered, and
that the    characteristics  of  this   slowing-down  (super-Arrhenius
activated temperature dependence of  the  primary relaxation time  and
non-exponential decay of the relaxation function) are similar to those
observed in fragile glassforming liquids\cite{grouss}.  As in liquids,
glassiness is self-generated and does not result from the introduction
of  quenched  spin-glass-like randomness or  of dynamical constraints.
An    important property  of  these models   is   that by varying  the
frustration    strength, i.    e.,   the   relative amplitude   of the
long-ranged frustrating interaction, one can  span the whole range  of
glassforming behavior,  from  strong (Arrhenius  T-dependence) to very
fragile  (marked super-Arrhenius T-dependence);  the less frustrated a
system, the more fragile it is.

\begin{figure}\vspace{-1cm}
\hspace{-1cm}
\resizebox{17cm}{!}{ \includegraphics{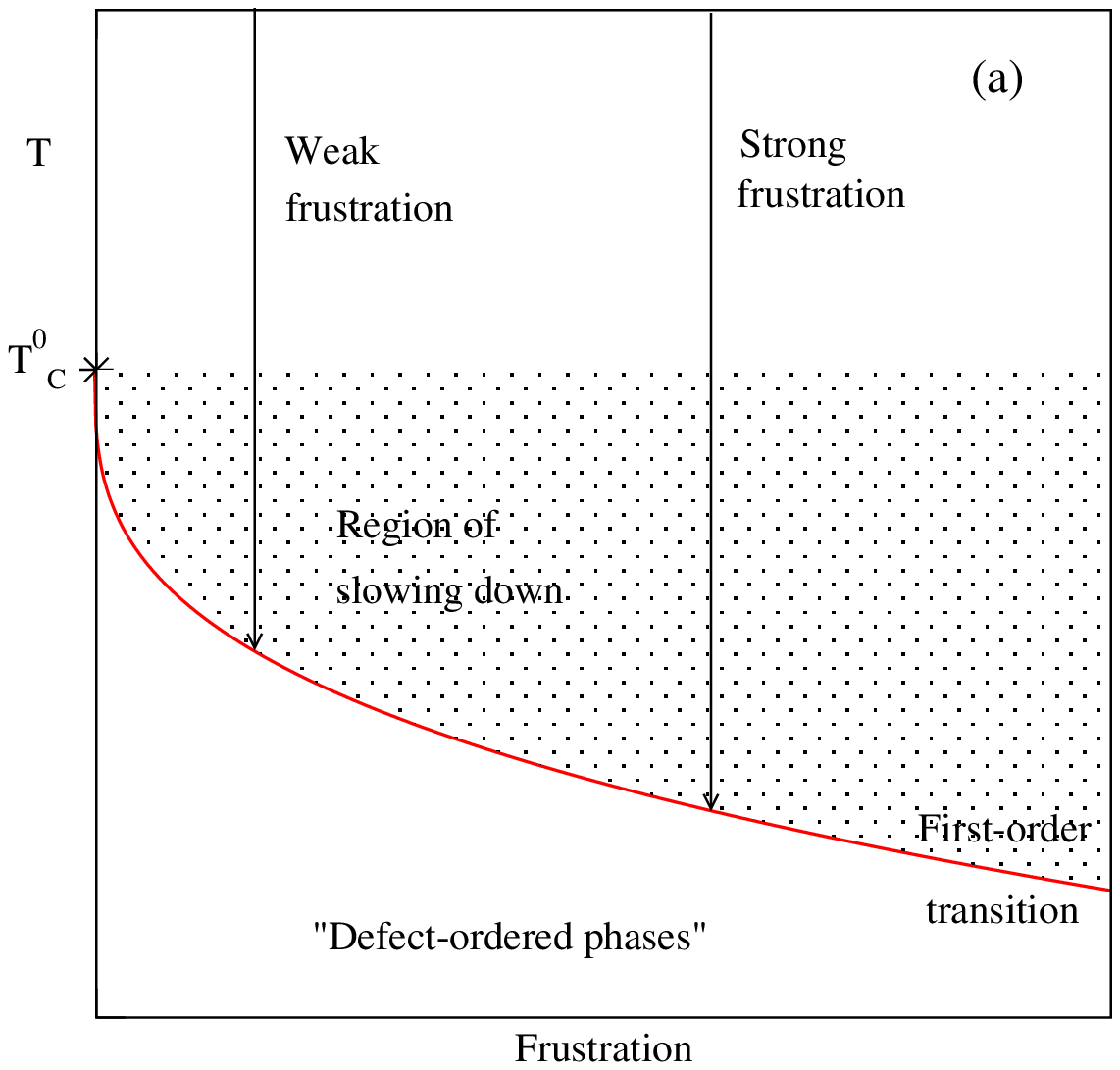}\hspace{-2cm}
\includegraphics{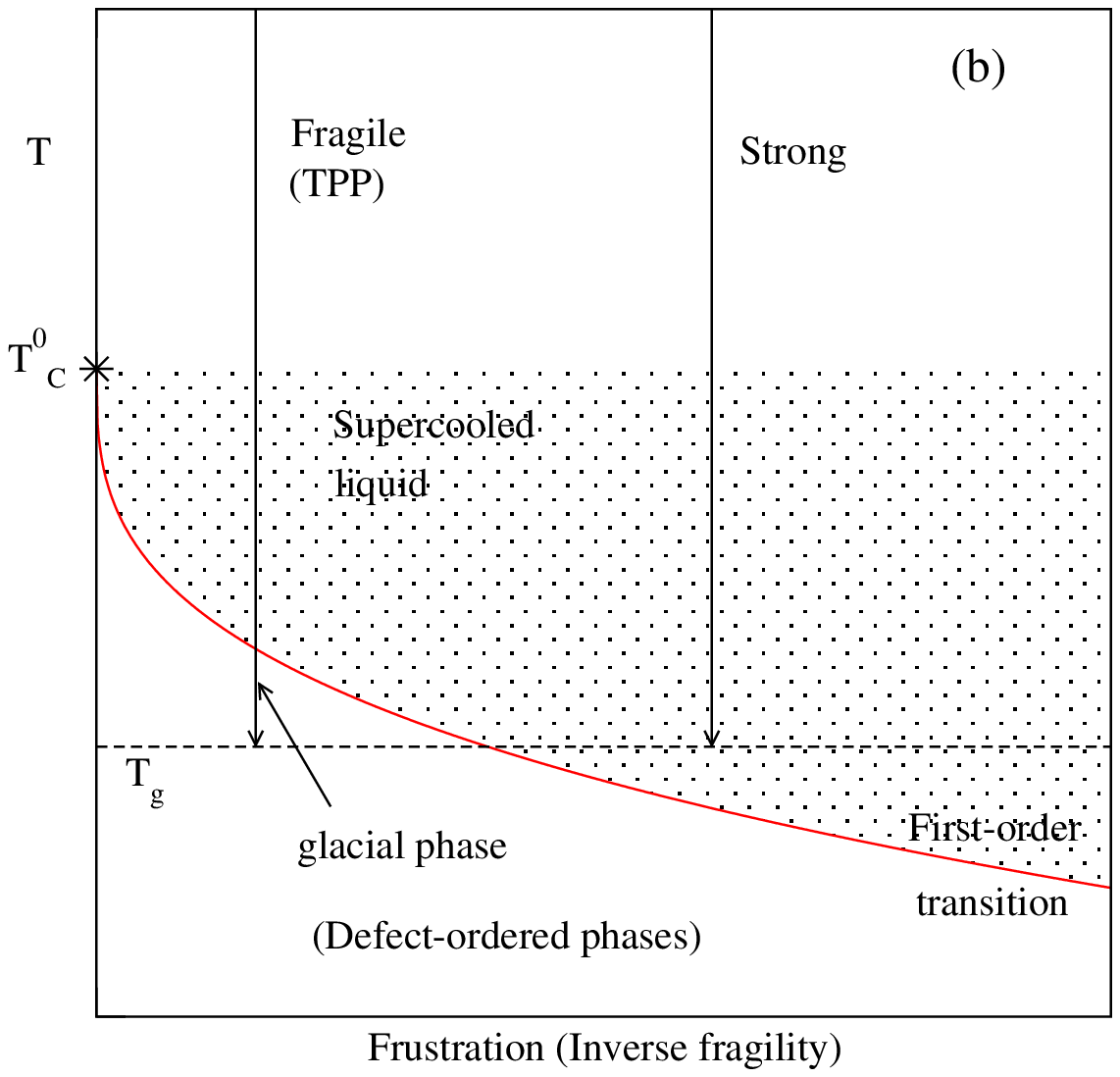}}
(c) \hspace*{2cm} weak frustration \hspace*{2cm}strong frustration

\begin{center}

\resizebox{11cm}{!}{\includegraphics{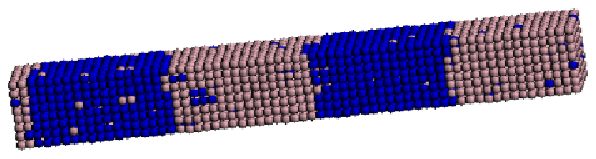}\includegraphics{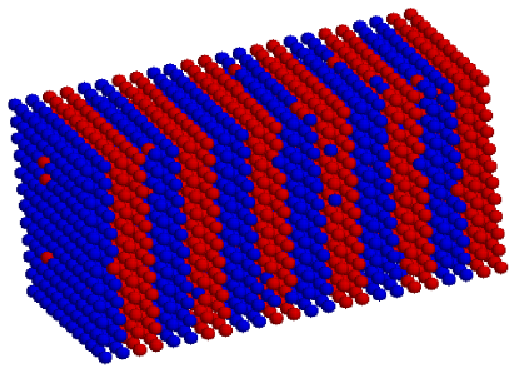}
}
\caption{a) Schematic temperature-frustration phase diagram of frustrated
 models with a long-range frustrating interaction.  The defect-ordered
 phases   are   illustrated   in  (c).  b)    Transposition     of the
 temperature-frustration phase   diagram to  a   temperature-fragility
 diagram  for glassforming liquids.    A liquid is characterized  by a
 given frustration and a given  fragility, the smaller the former, the
 larger  the   latter.  The  dotted   line represent  the hypothetical
 position of the   experimental glass transition temperature  that may
 come above   or   below the  transition  line  to  the defect-ordered
 phases.   c)    Low-temperature   configurations of the
 Coulomb frustrated Ising model obtained by Monte Carlo simulation for weak and strong frustration\cite{grouss}. }
\end{center}
\end{figure}
\section{Frustration and defect-ordered phases}
It has  been shown that, generically, frustration does indeed lead
to formation  of low-T defect-ordered phases  \cite{viot}. The situation is
schematically illustrated in Fig. 4a for  the Coulomb frustrated Ising
ferromagnet \cite{viot}. In the frustration-temperature diagram, there is a
line of first-order phase transition  from the high-T disordered phase
to the  low-T defect-ordered phases.  (Recall  that the usual crystal,
which is  more stable than the  liquid  and the  defect-ordered phases
below the melting point, is not included in  this picture.) The region
of  strong slowing-down is  above  the transition  line. For the Ising
model considered  here for illustration,  the defect-ordered phases at
low  frustration  are  lamellar   phases  whose period  increases   as
frustration decreases: see Fig.~4a. Said  differently, the size of the
unit  cell (here,  as in the   dodecagonal  quasi-crystal, there  is a
one-dimensional   periodic pattern and  ideal order  in the transverse
directions) increases as frustration decreases, and thus, as discussed
above, as  fragility increases.  The  details, e. g.,  the microscopic
characterization of the  ideal order, may  be model-dependent, but the
overall trends are robust.
	
Transposing the above results  to glassforming liquids, a liquid being
characterized by a given value   of the frustration strength, one  may
speculate that fragile     glassformers would  tend to    form   low-T
defect-ordered  phases  with  large unit   cells: see Fig.~4b.   It is
possible that due  to their large unit  cells  and the  fact that they
appear   in the viscous liquid  regime,   these phases would be poorly
crystallized, i.   e.,  appear as    powders  with small  polydisperse
crystallites when formed upon  decreasing the temperature.  This leads
us  to  suggest  that the  ``apparently  amorphous''   glacial phase  of
fragile TPP is a frustration-induced defect-ordered phase with a large
unit cell. Note that the possibility  of observing such phases depends
on  a non-universal property,  the relative position of the transition
temperature $T_{DO}$ with  respect to the glass transition temperature
$T_g$. Only if $T_g$ is less than $T_{DO}$  can a defect-ordered phase
be experimentally obtained, and this may be a rare situation.

\section{Conclusion}
	
We    have proposed  that    characterizing  liquids by their  locally
preferred  structure  (LPS) could help  in  understanding the observed
phenomenon of polyamorphism.  ``True polyamorphism'' would involve the
competition between  two (or more) distinct LPS's,  one favored at low
pressure  because of its low energy  and one favored  at high pressure
because of its small specific volume,  as in tetrahedrally coordinated
systems.  ``Apparent  polyamorphism'' that     we associate  with  the
existence of a poorly  crystallized defect-ordered phase with  a large
unit   cell and  small  crystallites,   could  result from  structural
frustration, i.  e., a competition  between the tendency to extend the
short-ranged  LPS and  a long-ranged  global  constraint that prevents
tiling  of  the whole  space   by the  LPS.  The  fragile  glassformer
triphenyl phosphite,  in  which a  first-order transition  is observed
between  the  supercooled   liquid  and the  mesoscopically structured
glacial  phase, may be  one  example of such ``apparent polyamorphism''.
Although some of these ideas have  been previously considered, this is
the first time that  they have been  incorporated within a  consistent
picture.

Finally,  it  is tempting   to     speculate that   the    low-density
``amorphous'' phase of water, a phase that shows none of the canonical
low-T  features of truly amorphous glasses   (excess in the density of
states at   low   frequency over  the  value   expected  from harmonic
vibrations, and violation of the Debye $T^3$ law due to the presence of
low-energy  two-level  systems)  \cite{ang2},   could  also   be  only
apparently amorphous, thereby adding more to the ``puzzling behavior of
water at very low temperature''\cite{stan}.

\end{document}